\documentclass[english,aps,twocolumn]{revtex4-1}
\usepackage[T1]{fontenc}
\usepackage[latin9]{inputenc}
\usepackage{amsmath}
\usepackage{amssymb}
\usepackage{graphicx}
\usepackage{esint}
\usepackage{babel}
\begin{document}

\title{Efficient simulations with electronic open boundaries}

\author{Andrew P. Horsfield$^{1}$, Max Boleininger$^{2}$, Roberto D'Agosta$^{3,4}$,
Vyas Iyer$^{1}$, Aaron Thong$^{1}$, Tchavdar N. Todorov$^{5}$,
Catherine White$^{1}$.}

\affiliation{$^{1}$Department of Materials and Thomas Young Centre, Imperial
College London, South Kensington Campus, London SW7 2AZ, U.K.}

\affiliation{$^{2}$Department of Physics and Thomas Young Centre, Imperial
College London, South Kensington Campus, London SW7 2AZ, U.K.}

\affiliation{$^{3}$Nano-Bio Spectroscopy Group and ETSF, Universidad del País
Vasco, CFM CSIC-UPV/EHU , 20018 San Sebastián, Spain}

\affiliation{$^{4}$IKERBASQUE, Basque Foundation for Science, E-48013 Bilbao,
Spain}

\affiliation{$^{5}$Atomistic Simulation Centre, School of Mathematics and Physics,
Queen\textquoteright s University Belfast, Belfast BT7 1NN, U.K.}

\email{a.horsfield@imperial.ac.uk}

\begin{abstract}
We present a reformulation of the Hairy Probe method for introducing
electronic open boundaries that is appropriate for steady state calculations
involving non-orthogonal atomic basis sets. As a check on the correctness
of the method we investigate a perfect atomic wire of Cu atoms, and
a perfect non-orthogonal chain of H atoms. For both atom chains we
find that the conductance has a value of exactly one quantum unit,
and that this is rather insensitive to the strength of coupling of
the probes to the system, provided values of the coupling are of the
same order as the mean inter-level spacing of the system without probes.
For the Cu atom chain we find in addition that away from the regions
with probes attached, the potential in the wire is uniform, while
within them it follows a predicted exponential variation with position.
We then apply the method to an initial investigation of the suitability
of graphene as a contact material for molecular electronics. We perform
calculations on a carbon nanoribbon to determine the correct coupling
strength of the probes to the graphene, and obtain a conductance of
about two quantum units corresponding to two bands crossing the Fermi
surface. We then compute the current through a benzene molecule attached
to two graphene contacts and find only a very weak current because
of the disruption of the $\pi$-conjugation by the covalent bond between
the benzene and the graphene. In all cases we find that very strong
or weak probe couplings suppress the current.
\end{abstract}


\maketitle

\section{Introduction}

Atomic scale computer simulations of nanoscale systems of necessity
have to approximate the environment that the system finds itself in
as it is of unlimited size. One way to incorporate a model environment
is through the boundary conditions of the system being treated explicitly.
Here we focus on the boundary conditions for the electrons. Traditional
choices include: free boundaries, where the system is treated as an
isolated cluster in vacuum; periodic boundaries, where the system
plus its near environment are repeated periodically to make an effectively
infinite system; and open boundaries, where the system is finite but
electrons can enter and leave as though connected to an external reservoir.
The correct choice of boundary conditions is determined by the problem
being addressed.

There exist very efficient algorithms for free and periodic boundary
atomistic simulations \cite{Martin2004}, and these will not be considered
further here. Open boundaries are important for a number of problems
\cite{Metzger2015}, and mature open boundary codes also exist \cite{Brandbyge2002,Rocha2004,Ozaki2010,ATK}.
However, relative to free and periodic boundaries, they tend to be
more computationally expensive to implement, and simulations can require
more human effort to set up. These technical considerations tend to
limit the range of problems addressed, often to molecular conduction,
whereas if they could be overcome new areas would become accessible,
such as electrochemistry. Our purpose here is to map out a possible
way forward by extending a light weight scattering theory scheme known
as Hairy Probes \cite{McEniry2007a} to systems more general than
those to which it was originally applied, and to show that simulations
can be made computationally efficient and easy to set up. Hairy Probes
originally was designed to address time dependent problems; here we
only consider the case of steady state current and static atoms.

Using an Empirical Tight Binding (ETB) model we investigate a Cu atomic
wire, and then using a Density Functional Tight Binding (DFTB) model
\cite{Frauenheim2000}, we apply the method to the study of a chain
of H atoms. These two simple, but well understood systems, allow us
to investigate the correctness of the method. For both the Cu and
H wires we get ideal ballistic conductance provided the strength of
the coupling to the probes is neither too large nor too small: extreme
couplings suppress the current. We then look at current flow through
a benzene molecule between two graphene contacts as a way to investigate
the properties of graphene as a contact for molecular electronics
\cite{AllenTungKaner2010,WangKimChoeEtAl2011}. We find that the presence
of a covalent bond between the benzene and the graphene suppresses
the current as it disrupts the $\pi$-conjugation. As preparation
for this calculation, current flow through a carbon nanoribbon is
studied to find the correct coupling strength for the probes to the
graphene. We obtain a conductance of slightly less than two quantum
units corresponding to two bands crossing the Fermi level.

\section{Formalism}

The Hairy Probes formalism was originally introduced for orthogonal
tight binding models, and covered both static and time dependent simulations
\cite{McEniry2007a}. Here we generalize the static limit to the non-orthogonal
case \cite{EmberlyKirczenow1998}, summarizing the key steps in the theory. The expressions are
derived using the Lippmann-Schwinger formalism \cite{Mujica1994,Todorov1993},
which is equivalent to using non-equilibrium Green's functions (NEGF)
for non-interacting or mean field Hamiltonians \cite{Wang2009}.

We note that this method has a number of similarities with the sink-source
potential method \cite{Goyer2007,Rocheleau2012,Pickup2015}; however,
additional simplifications allow for arbitrary bias, any number of
terminals, and full self-consistency. Similar simplifications have also
been achieved previously by applying the wide band limit directly to the
leads \cite{Verzijl2013}. However, we note that Hairy Probes can deliver accurate
results for low dimensional systems, and charge self-consistency can be introduced
straightforwardly, as demonstrated below.

The starting point is to imagine that our system is connected to one
or more particle reservoirs by a set of atomically thin leads (which
we call probes) that each attach to just one atomic orbital in the
system of interest. The reservoir of electrons from which a probe
emerges is characterized by a chemical potential and a temperature
for the electrons. Each probe, then, is a bit like a wire attached
at one end to a terminal of a battery, and at the other end attached
by a kind of alligator clip to an atomic orbital. Each probe thus
corresponds to both a source of incoming electrons of given chemical
potential and temperature, and a channel for outgoing electrons.

\begin{figure}[h]
\begin{centering}
\includegraphics[width=0.9\columnwidth]{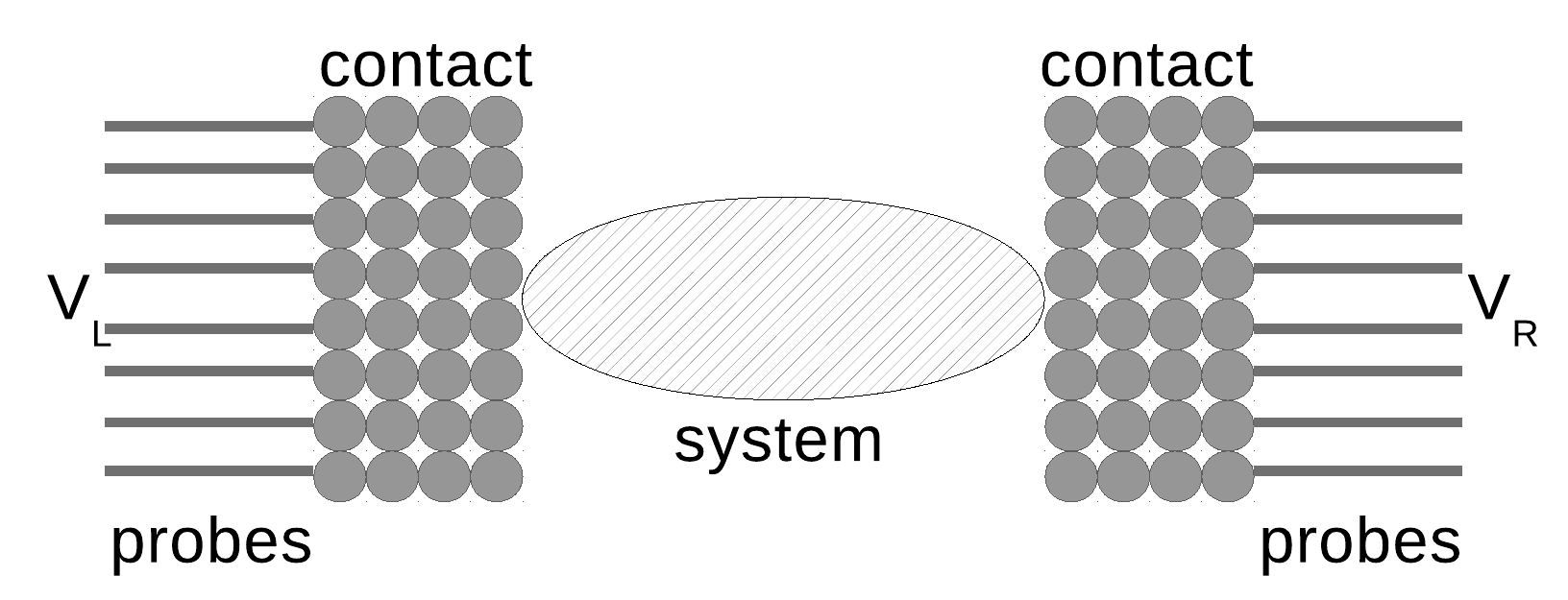}
\par\end{centering}
\caption{\label{fig:probes}The arrangement for a calculation involving
  hairy probes.}
\end{figure}

In practice, the probes are attached to contact regions in much the
same way that leads are attached to contacts in many NEGF calculations:
see Fig. \ref{fig:probes}. However, because the probes are not system
specific, we can define them in a manner that is computationally convenient.
Thus there is no need to compute surface Green's functions, the embedding
self-energy can be made energy independent while avoiding imposing
the wide band limit directly to the leads, and the mean field self-consistent
potential profile is taken care of automatically. These attributes
are what enable the Hairy Probe formalism to be easy to use (you just
need to specify where the probes are to be attached, and how strongly,
but do not have to build Green's functions for the leads), and computationally
very efficient (an effective Hamiltonian is produced that can be diagonalized,
and all subsequent integrals can then be performed analytically).
We note that it is shown in \cite{McEniry2007a} that in the limit
of long electrodes and small coupling the Hairy Probes steady state
reduces to the conventional 2-terminal Landauer picture.

The argument we present here that leads to the Hairy Probes equations
is based on the Lippmann-Schwinger formulation of scattering theory.
That is, we treat each probe as transporting independent electrons
from a reservoir, with the electron wavefunctions being viewed as
scattering states that travel down the probes and scatter off the
system of interest, being partially transmitted (producing a current)
and partially reflected.

As we employ ETB and DFTB models, the basis set used to expand the
single particle wave functions is composed of atomic orbitals. Let
our atomic basis set be denoted by $\left|\alpha\right\rangle $ where
$\alpha$ is a combined index that spans both atomic sites and orbitals.
We can now define the Hamiltonian and overlap matrices by $H_{\alpha\alpha'}=\left\langle \alpha\left|\hat{H}\right|\alpha'\right\rangle $
and $S_{\alpha\alpha'}=\left\langle \alpha\mid\alpha'\right\rangle $,
respectively. Note that the Hamiltonian includes all the terms associated
with self-consistent charge redistribution \cite{Frauenheim2000,Horsfield2012}.
We partition these orbitals between the system ($\left|\beta\right\rangle $)
and the probes ($\left|p\gamma_{p}\right\rangle $), where $p$ is
the index of the probe and $\gamma_{p}$ indexes an orbital in probe
$p$.

Consider a state with index $n_{p}$ in probe $p$ with energy $E_{pn_{p}}$
that is stationary before the probe is connected to the system. Let
us denote this state by $\left|\phi^{(pn_{p})}\right\rangle =\sum_{\gamma_{p}}\phi_{\gamma_{p}}^{(pn_{p})}\left|p\gamma_{p}\right\rangle $
where $\phi_{\gamma_{p}}^{(pn_{p})}$ is an expansion coefficient.
A scattered wave forms from this state after the probe is attached
to the system, which we denote by $\left|\psi^{(pn_{p})}\right\rangle =\sum_{\beta}\psi_{\beta}^{(pn_{p})}\left|\beta\right\rangle +\sum_{p'\gamma_{p'}}\psi_{p'\gamma_{p'}}^{(pn_{p})}\left|p'\gamma_{p'}\right\rangle $,
where $\psi_{\beta}^{(pn_{p})}$ is an expansion coefficient for orbitals
in the system, and $\psi_{p'\gamma_{p'}}^{(pn_{p})}$ is an expansion
coefficient for orbitals in probe $p'$. The scattered wave is related
to the initial state by the Lippmann-Schwinger equation, giving
\begin{equation}
\psi_{\beta}^{(pn_{p})}=\sum_{\beta'\gamma_{p}}G_{\beta\beta'}^{R}(E_{pn_{p}})W_{\beta',p\gamma_{p}}(E_{pn_{p}})\phi_{\gamma_{p}}^{(pn_{p})}\label{eq:formalism-01}
\end{equation}
where $G^{R}$ is the retarded Green's function matrix for the whole
system, including all probes, and $W_{\beta',p\gamma_{p}}(E)=H_{\beta,p\gamma_{p}}-ES_{\beta,p\gamma_{p}}$
is an effective coupling matrix element between the system and probe
$p$. If we now define the retarded self energy $\Sigma_{\beta\beta'}^{(p)R}(E)=\sum_{\gamma_{p}\gamma_{p}'}W_{\beta,p\gamma_{p}}(E)G_{\gamma_{p},\gamma_{p}'}^{(p)R}(E)W_{p\gamma_{p}',\beta'}(E)$,
where $G_{\gamma_{p},\gamma_{p}'}^{(p)R}$ is the retarded Green's
function matrix for isolated probe $p$, we get the following central
results
\begin{eqnarray}
\delta_{\beta\beta''} & = &
                            \sum_{\beta'}\left(\left(E+\mathrm{i}\eta\right)S_{\beta\beta'}-H_{\beta\beta'}-\sum_{p}\Sigma_{\beta\beta'}^{(p)R}(E)\right)\nonumber\\
&&\times G_{\beta'\beta''}^{R}(E)\label{eq:formalism-02a}\\
\rho_{\beta\beta'} & = & \frac{1}{2\pi\mathrm{i}}\sum_{p}\int
                         f^{(p)}(E)\sum_{\beta''\beta'''}G_{\beta\beta''}^{R}(E)\nonumber\\
&&\times\left\{ \Sigma_{\beta''\beta'''}^{(p)A}(E)-\Sigma_{\beta''\beta'''}^{(p)R}(E)\right\} G_{\beta'''\beta'}^{A}(E)\,\mathrm{d}E\label{eq:formalism-02b}
\end{eqnarray}
where $\eta$ is a positive infinitesimal, $G_{\beta'''\beta'}^{A}$
and $\Sigma_{\beta''\beta'''}^{(p)A}$ are the advanced Green's function
and self energy respectively, $\rho_{\beta\beta'}$ is the single
particle electronic density matrix, and $f^{(p)}(E)$ is the occupancy
of the levels inside the isolated probe $p$, and hence the energy
distribution with which electrons are injected into the system by
probe $p$.

We now introduce the Hairy Probe anzatz for the retarded self energy.
We note that we want the simplest possible form that still possesses
the properties required by a self energy. Making it (almost) energy
independent allows us to reduce the problem of finding the scattering
states to a simple diagonalization, and by making it local to one
orbital we minimize the parameters we have to set. We then end up
with the following form 
\begin{equation}
\Sigma_{\beta\beta'}^{(p)R}(E)=\delta_{\beta\beta'}\begin{cases}
-\frac{1}{2}\mathrm{i}\Gamma_{p}\delta_{\beta\beta_{p}} & E\ge E_{p,c}\\
0 & E<E_{p,c}
\end{cases}\label{eq:formalism-03}
\end{equation}
where $\beta_{p}$ is the index of the orbital in the system to which
the probe $p$ is attached, and $E_{p,c}$ is the bottom of the band
for the electronic states in probe $p$, taken to be well below any
energy levels in the system. As the self energies are imaginary, they
have the effect of allowing electrons to be added to, and removed
from, the system \cite{Goyer2007}. The quantities $\Gamma_{p}$ set
the broadening of the states in the system, and define the rates at
which electrons can enter or leave.

Provided $E_{p,c}$ lies below all levels in the system, then we can
substitute Eq. \ref{eq:formalism-03} into Eqs. \ref{eq:formalism-02a}
and \ref{eq:formalism-02b} to give
\begin{eqnarray}
\delta_{\beta\beta''} & = &
                            \sum_{\beta'}\left(\left(E+\mathrm{i}\eta\right)S_{\beta\beta'}-\left[H_{\beta\beta'}-\delta_{\beta\beta'}\frac{\mathrm{i}}{2}\sum_{p}\Gamma_{p}\delta_{\beta\beta_{p}}\right]\right)\nonumber\\
&&\times G_{\beta'\beta''}(E)\label{eq:formalism-04a}\\
\rho_{\beta\beta'} & = & \frac{1}{2\pi}\sum_{p}\Gamma_{p}\int_{E_{p,c}}^{\infty}f^{(p)}(E)G_{\beta\beta_{p}}^{R}(E)G_{\beta_{p}\beta'}^{A}(E)\,\mathrm{d}E\label{eq:formalism-04b}
\end{eqnarray}
Note that Eq. \ref{eq:formalism-04b} offers an alternative, albeit
unphysical, interpretation of $E_{p,c}$: it is the energy of the
lowest occupied state in probe $p$, with states with $E<E_{p,c}$
being unoccupied. A discussion of the implications of the choice of
$E_{pc}$ is presented in the Appendix C.

We write the retarded Green's function as 
\begin{equation}
G_{\beta\beta'}^{R}(E)=\sum_{r}\frac{\chi_{\beta}^{(r)}\zeta_{\beta'}^{(r)*}}{E+\mathrm{i}\eta-\epsilon^{(r)}}\label{eq:formalism-05}
\end{equation}
where $\zeta_{\beta}^{(r)}$ and $\chi_{\beta'}^{(r)}$ are left
and right eigenstates, and $\epsilon^{(r)}$ the corresponding complex
eigenvalue. These satisfy 
\begin{eqnarray}
\sum_{\beta'}\left[H_{\beta\beta'}-\delta_{\beta\beta'}\sum_{p}\frac{\mathrm{i}}{2}\Gamma_{p}\delta_{\beta\beta_{p}}\right]\chi_{\beta'}^{(r)} & = & \epsilon^{(r)}\sum_{\beta'}S_{\beta\beta'}\chi_{\beta'}^{(r)}\label{eq:formalism-06a}\\
\delta_{rs} & = & \sum_{\beta\beta'}\zeta_{\beta}^{(r)*}S_{\beta\beta'}\chi_{\beta'}^{(s)}\label{eq:formalism-06b}
\end{eqnarray}
In principle setting $\chi_{\beta}^{(r)}=\zeta_{\beta}^{(r)*}$ should
satisfy Eq. \ref{eq:formalism-06b} as the Hamiltonian matrix is symmetric,
but we have found that better results are found by solving Eq. \ref{eq:formalism-06b}
explicitly, especially in the presence of degeneracies.

To solve these equations, the numerical procedure we have adopted is
as follows. We first transform Eq. \ref{eq:formalism-06a} from a
generalised eigenvalue problem to an ordinary one in the usual way. First we
carry out a Cholesky decomposition of the overlap matrix and use the
resulting triangular matrices to express the Hamiltonian (including
the self-energies) in an orthogonal representation. We then
diagonalise the Hamiltonian matrix using a general complex eigensolver
as the problem is complex and symmetric, rather than Hermitian, and
obtain the complex eigenvalues and right eigenvectors. The left
eigenvectors are then obtained by inverting the square matrix of right
eigenvectors, and then all eigenvectors are transformed back to the
original representation using the triangular matrices from the
Cholesky decomposition.

Substituting Eq. \ref{eq:formalism-05} into Eq. \ref{eq:formalism-04b}
gives 
\begin{equation}
\rho_{\beta\beta'}=\sum_{rs}f_{rs}\chi_{\beta}^{(r)}\chi_{\beta'}^{(s)*}\label{eq:formalism-06e}
\end{equation}
where 
\begin{equation}
f_{rs}=\frac{1}{2\pi}\sum_{p}\Gamma_{p}\zeta_{\beta_{p}}^{(r)*}\zeta_{\beta_{p}}^{(s)}\int_{E_{p,c}}^{\infty}\frac{f^{(p)}(E)}{\left(E-\epsilon^{(r)}\right)\left(E-\epsilon^{(s)*}\right)}\,\mathrm{d}E\label{eq:formalism-06c}
\end{equation}
and can be thought of as a generalized occupancy. We note that in
the limit of very weakly coupled leads ($\Gamma_{p}\to0$) all having
the same coupling strength, the occupancy simplifies to
\begin{equation}
f_{rs}\to\delta_{rs}\frac{\sum_{p}f^{(p)}(\Re\epsilon^{(r)})|\zeta_{\beta_{p}}^{(r)}|^{2}}{\sum_{p}|\zeta_{\beta_{p}}^{(r)}|^{2}}\label{eq:frs-small-G}
\end{equation}
This limiting form of the occupancy matrix is real and diagonal, so
the system carries no current, and is a weighted sum of the contributions
from each probe. We include it in the spirit of moving open boundaries
to problems outside the usual range, as it might be relevant to the
case of an electrode in an electrochemical cell. Finally we note that
if the populations $f^{(p)}(\epsilon)$ are independent of the probes,
then we get back the usual equilibrium expression for the density
matrix.

We compute the current through the bond between orbitals $\beta$
and $\beta'$ from 
\begin{equation}
\mathcal{I}_{\beta\beta'}=-\frac{4e}{\hbar}\left(H_{\beta\beta'}\mathrm{Im}\rho_{\beta\beta'}-S_{\beta\beta'}\mathrm{Im}E_{\beta\beta'}\right)\label{eq:formalism-06d}
\end{equation}
where $E_{\beta\beta'}=\sum_{rs}g_{rs}\chi_{\beta}^{(r)}\chi_{\beta'}^{(s)*}$
and $g_{rs}=\frac{1}{2\pi}\sum_{p}\Gamma_{p}\zeta_{\beta_{p}}^{(r)*}\zeta_{\beta_{p}}^{(s)}\int_{E_{p,c}}^{\infty}\frac{Ef^{(p)}(E)}{\left(E-\epsilon^{(r)}\right)\left(E-\epsilon^{(s)*}\right)}\,\mathrm{d}E$,
and a factor of 2 for spin degeneracy has been included. A derivation
of this expression is given in the Appendix B. We use finite temperature
occupations for the electrons in the probes. To enable analytic and
efficient evaluation of the integrals involving the occupancies, we
use the following piecewise linear approximation to the Fermi-Dirac
distribution function:
\begin{equation}
f^{(p)}(E)=\begin{cases}
1 & E-\mu_{p}\le-2k_{B}T_{p}\\
\frac{1}{2}-\frac{1}{4}\left(\frac{E-\mu_{p}}{k_{B}T_{p}}\right) & -2k_{B}T_{p}<E-\mu_{p}<+2k_{B}T_{p}\\
0 & E-\mu_{p}\ge+2k_{B}T_{p}
\end{cases}\label{eq:formalism-07}
\end{equation}
where $\mu_{p}$ and $T_{p}$ are the chemical potential and temperature
for the electrons in probe $p$. The integrals are given in the Appendix A.

Finally we note that the transmission between two probes $p_1$ and
$p_2$ is given by
\begin{equation}
T_{12}(E)=\left|G^R_{\beta_{p_{2}}\beta_{p_{1}}}(E)\right|^{2}\Gamma_{p_{1}}\Gamma_{p_{2}.}\label{eq:transmission}
\end{equation}

\section{Results}

\subsection{Atomic wires}

\begin{figure}[h]
\begin{centering}
\includegraphics[width=0.9\columnwidth]{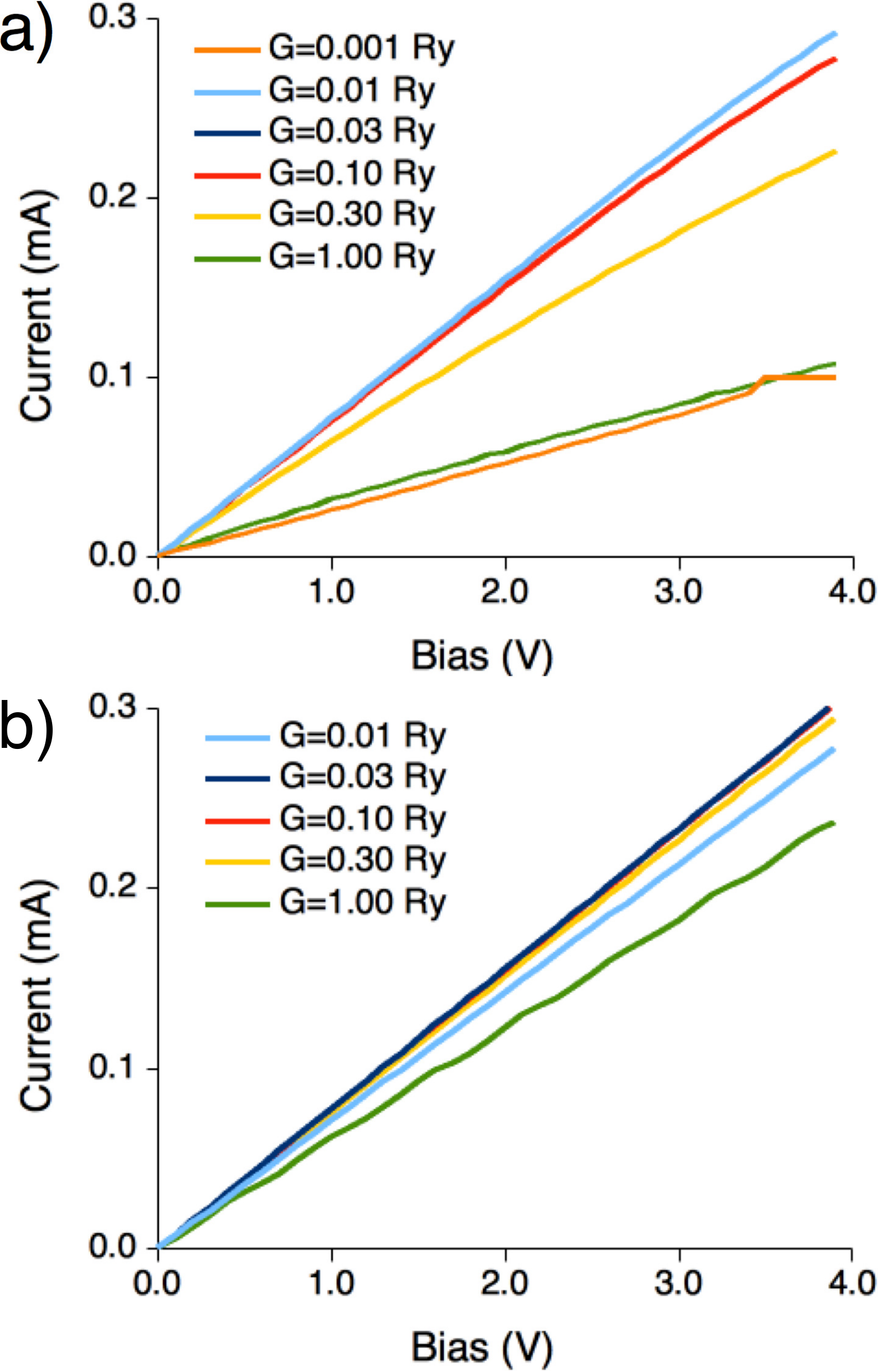}
\par\end{centering}
\caption{\label{fig:wire-IV}(Color online) a) The current through a wire composed of 300
Cu atoms as a function of the applied bias for a range of probe coupling
strengths (indicated by the symbol G). b) The current through a chain
of 300 H atoms as a function of the bias voltage for a range of probe
coupling strengths (indicated by the symbol G).
Note that in both panels the curves for coupling strengths of 0.1 Ry
and 0.03 Ry lie on top of each other.
}
\end{figure}

\begin{figure}
\begin{centering}
\includegraphics[width=0.9\columnwidth]{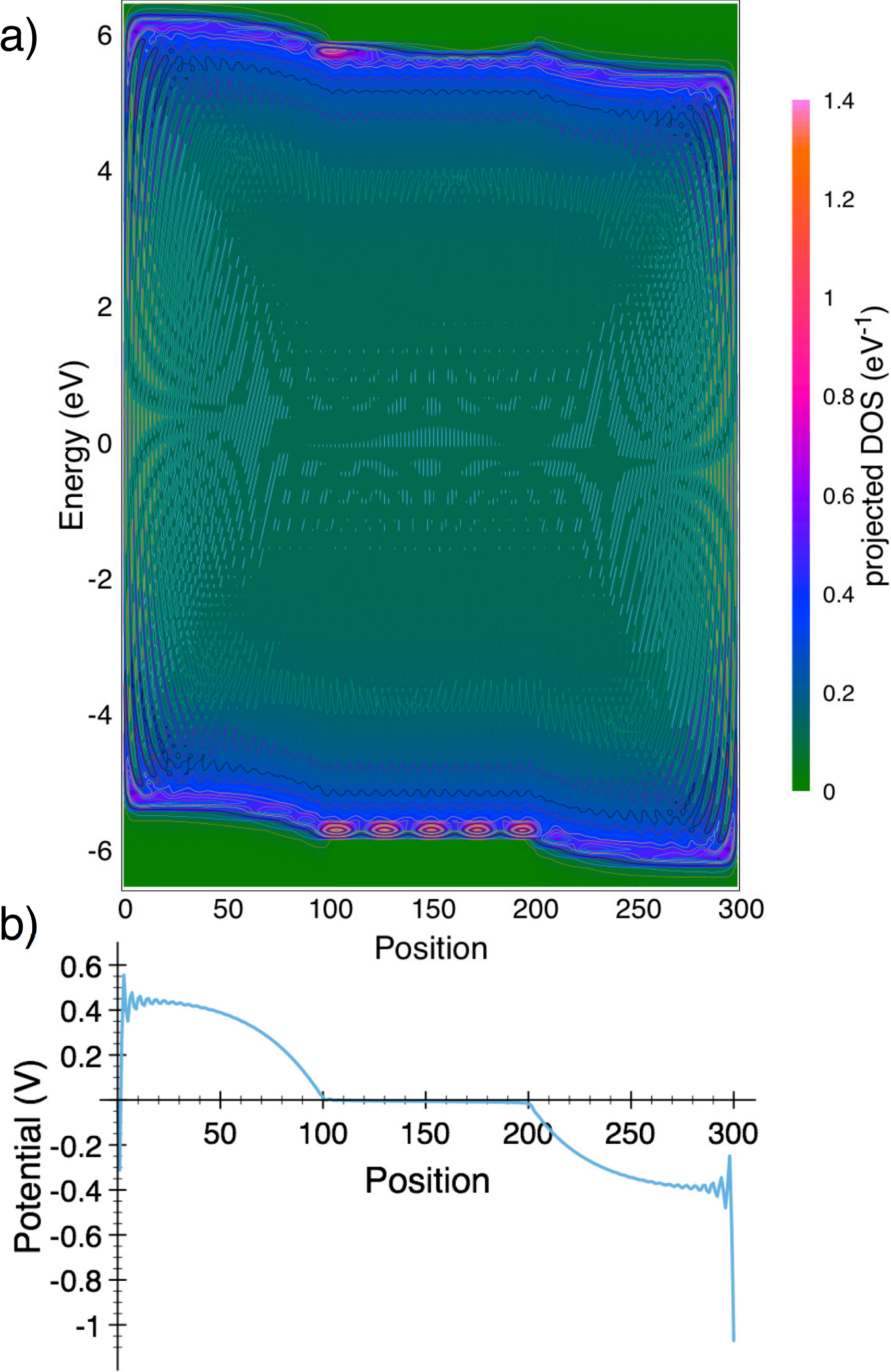}
\par\end{centering}
\caption{\label{fig:wire-dos} (Color online) a) A contour plot of the density of states projected
onto each atom. The position corresponds to the index of each atom
in the wire. The green regions correspond to low density of states,
while the blue and red regions correspond to a high density of states.
b) The average potential as a function of position.}
\end{figure}

The Hairy Probe algorithm has been implemented in the tight binding
program Plato \cite{Kenny2009}. To test the method we first investigated
an atomic wire made from 300 Cu atoms; probes were attached to the
first 100 and last 100 atoms. We used the orthogonal TB parameterization
of Sutton \emph{et al.} \cite{Sutton2001} that assigns just one s
orbital to each atom.
That is, there are 200 probes in all, one per orbital on each of the
200 lead atoms.
The probes all have the same coupling strength
$\Gamma_{p}$, and the same temperature $k_{B}T_{p}=0.001$ Ry. Open
boundary calculations are carried out in two stages. First, every
probe is assigned the same chemical potential, and its value is adjusted
until the system as a whole is charge neutral; this we term the reference
chemical potential. Each atom individually is allowed to acquire a
net charge, described by a monopole with a gaussian charge
distribution \cite{Soin2011},
and charge self-consistency is imposed. Second, a bias is applied
with the chemical potential on the left probes being raised by half
the bias relative to the reference chemical potential, and the chemical
potential on the right probes being lowered by half the bias. This
allows the wire to acquire a net charge, though this is typically
less than 1 electron for the whole system for biases up to 3.9 V.
The first step is necessary because the probes do not correspond to
a known physical system, so an anzatz is needed to give them sensible
characteristics.

We computed the current as a function of applied voltage for a range
of coupling strengths of the probes; the results are shown in Fig.
\ref{fig:wire-IV} a). We see that the current varies close to linearly
with bias for all probe coupling strengths, and that the slope (conductance)
is nearly independent of that coupling for values in the range 0.01
Ry to 0.10 Ry, and in this range the slope is equal to the quantum
unit of conductance ($G_{0}=7.748\times10^{-5}$ S). The current is
reduced for both larger and smaller couplings. With small couplings
the current is restricted by the rate at which charge can be injected
and removed by the probes. At very large couplings the hopping matrix
elements between atoms in the wire become a weak perturbation on the
interaction between the atoms and the probes; in this limit incoming
electrons are reflected back into the probes before they can contribute
to the current in the wire. The lower and upper bounds for reasonable
couplings are roughly the mean spacing between levels (to ensure we
have a continuous density of states) and the bandwidth (to ensure
the probes do not overwhelm the system).

In Fig. \ref{fig:wire-dos} a) is shown the density of states (DOS)
projected onto each atom (the atom index is on the x axis) as a function
of the electron energy (y axis) for a wire with a bias of 1 V applied,
and a coupling of 0.01 Ry for each probe. We see that in the middle
of the wire (atom position 150) we have a DOS that is sharply peaked
at the band edges. This is consistent with the cosine band structure
associated with an infinite chain of atoms with one s orbital per
atom. At the ends of the wire there is considerable weight towards
the middle of the energy range, consistent with the square root type
DOS associated with the end atom of a semi-infinite chain of s orbitals.
We note that the states in the lead regions (atoms 1 to 100, and 201
to 300) are significantly broadened by the probes.
Finally, the potential in the wire region is essentially independent
of position (Fig. \ref{eq:formalism-03} b)). This is to be contrasted
with the interface regions where the probes end and begin; here there
is a clear variation of potential with position suggesting that this
is where the potential drop occurs.

The variation of potential with position can be understood in the
following way. The potential in the probe free wire is uniform as
it is metallic and the electrons can move to screen out any charge
accumulation; current in a perfect conductor requires no field, locally
\cite{Landauer1989}. That leaves the regions with probes. Consider
electrons arriving at the left region with probes from the middle
region, with energies within the conduction window. In this region,
the lifetime of electrons before being absorbed into a probe is $\tau=\hbar/\Gamma$
and $\lambda=v_{group}\tau$ is the mean free path, with $v_{group}$
being the group velocity of the electrons at the Fermi energy. For
a cosine band with band filling $\xi$ we have $v_{group}=2 a v \sin(\pi\xi)/\hbar$,
where $a$ is the interatomic spacing, and $v$ is the hopping integral between neighbouring sites. The
fraction of electrons that make it to position $x$ (measured from
the junction between the perfect wire and the region with probes)
dies out as $\exp(-x/\lambda)$. To keep the metal neutral, the band-bottom
has to adopt the same shape, to compensate. We thus have the following
form for the potential at position $x$

\begin{equation}
\phi(x)\sim\frac{1}{2}eV\left(\exp\left(-\frac{x}{\lambda}\right)-1\right)\label{eq:phi-01}
\end{equation}
where $V$ is the applied voltage.

The functional form of $\phi(x)$ clearly has a shape corresponding
to that seen in Fig. \ref{fig:wire-dos} b). From Eq. \ref{eq:phi-01}
we get $\phi(x)/\phi(\infty)=1-\exp(-x/\lambda)$. If we let $x_{1/2}$
be the point where $\phi(x_{1/2})/\phi(\infty)=\frac{1}{2}$ then
we get $\lambda=x_{1/2}/\ln2$. From Fig. \ref{fig:wire-dos} b) we
see that $x_{1/2} \approx 20 a$ and hence $\lambda \approx 29 a$. As the
hopping integral is $v=0.212$ Ry, the band filling is 0.243 \cite{Sutton2001},
and $\Gamma_{p}=0.01$ Ry, we would expect $\lambda/a = 2v\sin(\pi\xi)/\Gamma_{p} \approx 29$;
this is in full agreement with the measured value.

We have repeated the above calculations using a non-orthogonal DFTB
model for hydrogen \cite{PorezagFrauenheimKoehlerEtAl1995}: an atomic
wire made from 300 H atoms with probes attached to the first 100 and
last 100 atoms. The resulting current against bias plot is shown in
Fig. \ref{fig:wire-IV} b). We see that it has the same structure
as for the orthogonal Cu wire (see Fig. \ref{fig:wire-IV} a)), and
that the maximum conductance is again one quantum unit. This suggests
that the method for including overlap into the formalism is correct.

We note that, for the case of orthogonal tight binding, agreement with
the two terminal Landauer solution was demonstrated previously for a
non-uniform wire, provided a sufficiently large number of probes was
employed\cite{McEniry2007a}. 

\subsection{Graphene contacts}

\begin{figure}[h]
\begin{centering}
\includegraphics[width=0.9\columnwidth]{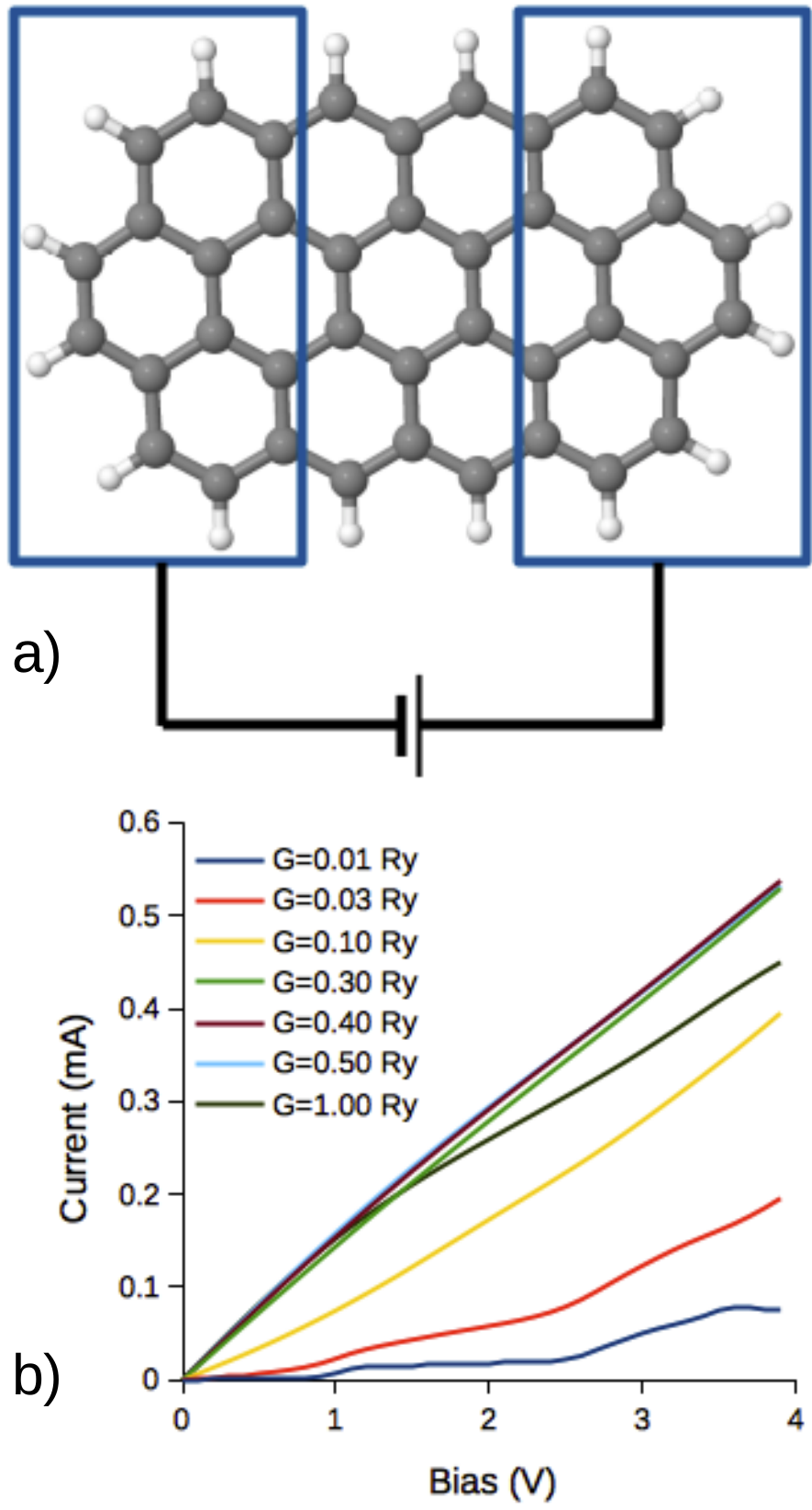}
\par\end{centering}
\caption{\label{fig:nanoribbon} (Color online) a) The arrangement for the Hairy Probe calculation
of current through a nanoribbon. The probes are attached to the atoms
within the blue boxes. b) The current through the nanoribbon as a
function of the bias voltage for a range of probe coupling strengths
(indicated by the symbol G).}
\end{figure}

Having studied simple one dimensional atomic wires, and found good
agreement with the expected conductance, we now consider electron
transport through a more complex system: a benzene ring attached to
two graphene contacts by means of covalent bonds. We have selected
this system because graphene's electrical properties \cite{AllenTungKaner2010}
suggest it might make a good contact material for molecular
electronics \cite{WangKimChoeEtAl2011}.
As we shall see below, care will have to be taken with how connection
to the contacts is made. We note that this system has some similarities
to the well studied benzene-dithiol between two gold contacts \cite{Delaney2004}.

To estimate the correct coupling strength of the probes to the graphene
contacts we first perform calculations of current through a carbon
nanoribbon. To compute the current through a small carbon nanoribbon,
whose edges have been terminated with hydrogen atoms (see Fig. \ref{fig:nanoribbon}),
we again use a non-orthogonal DFTB model \cite{PorezagFrauenheimKoehlerEtAl1995}.
The probes all have the same coupling strength $\Gamma_{p}$, and
same temperature $k_{B}T_{p}=0.001$ Ry. The variation of current
with bias is shown in Fig. \ref{fig:nanoribbon} for a range of coupling
strengths. For coupling strengths of 0.1 Ry and below we find that
the current increases roughly linearly with coupling strength for
a given bias, and is sensitive to details of the electronic structure
of the nanoribbon. The current is fairly insensitive to coupling strength
for $0.3\,\mathrm{Ry}\le\Gamma_{p}\le1\,\mathrm{Ry}$. At large coupling
strengths the current is again heavily suppressed. From this we conclude
that for carbon flakes of this size, setting $\Gamma_{p}=0.4$ Ry
is appropriate. At this coupling, the conductance is $1.44\times10^{-4}$
S which is 1.9 times the quantum of conductance; this can be understood
as resulting from two bands crossing the Fermi energy forming two
conductance channels.

\begin{figure}
\begin{centering}
\includegraphics[width=0.9\columnwidth]{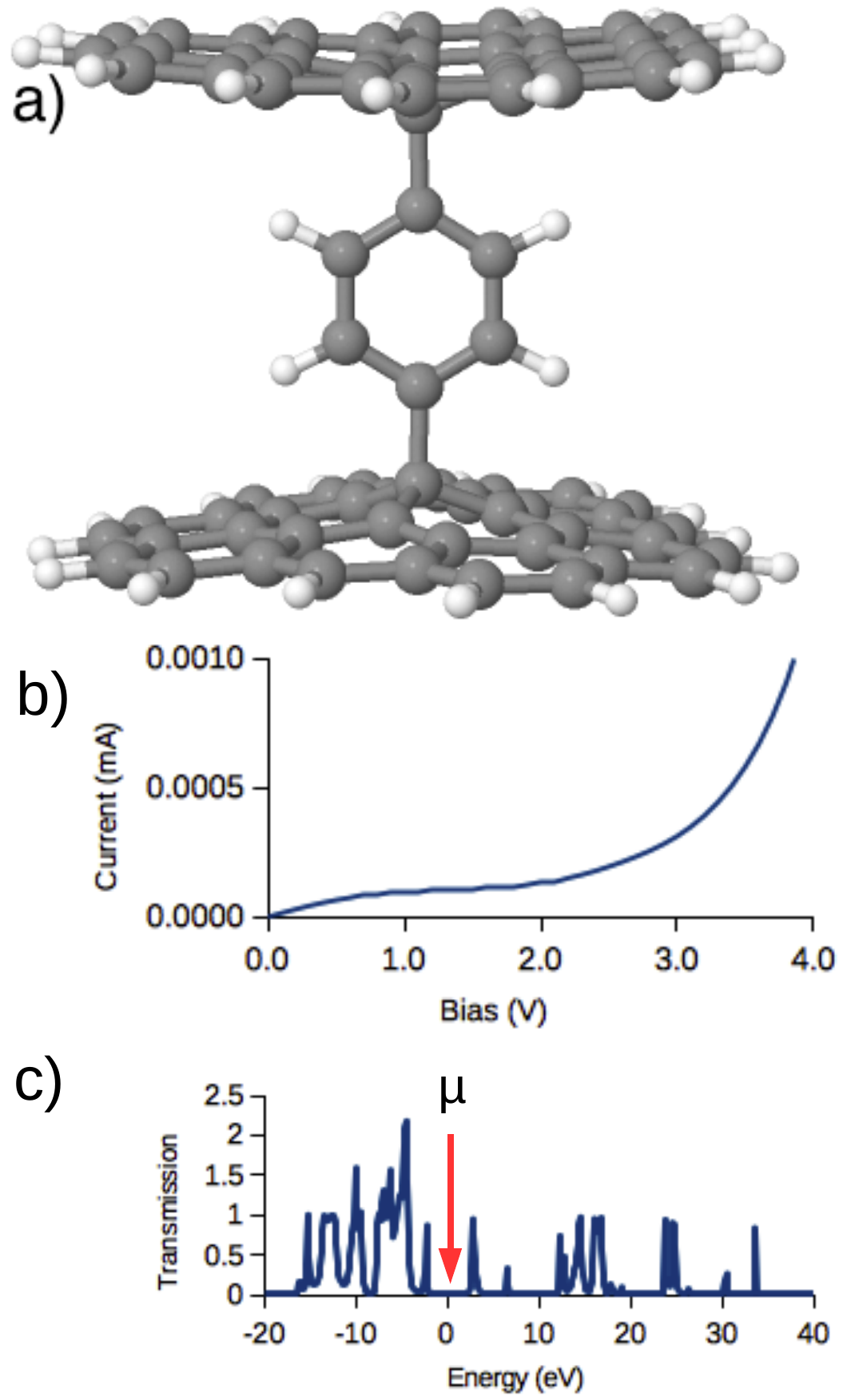}
\par\end{centering}
\caption{\label{fig:Graphene-Benzene-Graphene}(Color online) a) The
arrangement of the atoms for the calculation of a current through a
benzene molecule attached
covalently to two graphene flakes. The probes are attached to the
edges of the two graphene flakes. b) The current through the benzene
molecule as a function of the bias applied between the two graphene
contacts.
c) The transmission through the contacts and benzene molecule. The red
arrow indicates the location of the chemical potential of the probes at zero bias.
}
\end{figure}

Our final simulation is now of the current through a benzene ring
coupled covalently to a pair of graphene contacts. The contacts are
modelled as small flakes, whose edges are terminated with hydrogen
(see Fig. \ref{fig:Graphene-Benzene-Graphene} a)). The probes are
then attached to the atoms around the edges of each flake, with the
probes on one flake all having the same electron chemical potential.
The difference between the potentials of the two flakes then creates
the bias across the benzene molecule. We use the probe coupling strength
of $\Gamma_{p}=0.4$ Ry found from our nanoribbon calculations. Comparing
the current versus voltage plot from Fig. \ref{fig:Graphene-Benzene-Graphene}
b) with that from Fig. \ref{fig:nanoribbon} b), the first thing to
notice is that the current has dropped by a factor of over 1000. This
can be understood by looking at the transmission function for the
the benzene molecule (Fig \ref{fig:Graphene-Benzene-Graphene} c)).
Here we see that the reference chemical potential sits well within
a tunelling gap several eV wide, thus there are very few free carriers.
As the bias increases a small number of holes appear in the valence
band; the benzene molecule acquires a small positive charge of order
0.03$e$, which grows between 1.5V and 4V to about 0.04$e$. The presence
of the band gap is a consequence of the covalent bond between the
benzene ring and the graphene: at the point of contact, the carbon
atom in the graphene adopts sp$^{3}$ hybridization, disrupting the
$\pi$-conjugation. Thus, to form a good contact, a method is required
that maintains the conjugation. Finally, we note that the transport
is dominated by holes rather than electrons because the reference
chemical potential lies about 0.46 eV closer to the valence band than
to the conduction band.

\subsection{Graded probes}

Above we have applied the simplest implementation of the Hairy
Probes battery, where all probes have the same coupling strength to their
respective atoms. This implementation has the conceptual advantage of
corresponding most closely to the physical interpretation of the Hairy 
Probes as external particle baths, in which the system is immersed.
In Ref. \cite{McEniry2007a} it was shown that when the length of the hairy 
leads increases, and $\Gamma$ decreases (while always remaining larger 
than the lead energy-level spacing), the Hairy Probes steady state tends 
to the conventional 2-terminal Landauer steady state.

However in practice one would like to keep the leads as short as
possible for computational reasons. The rough rule of thumb for the optimal 
$\Gamma$ then is that it should be as small as possible, while remaining larger
than the level spacing in the leads. The resultant steady states then
approximate the conventional 2-terminal limit, but not exactly.
This is not right or wrong as the Hairy Probe battery is intended 
to be a stand-alone transport setup, with its own interpretation (as above). 
But the need to consider finite-size effects, and the precise choice of $\Gamma$, 
could then be seen as irksome.

To overcome this complication, a simple alternative is to make $\Gamma$
position-dependent, so that its value rises gradually from zero, as we move
along each lead, away from the central region. We refer to this scenario
as Graded Probes. Below we compare these two implementations numerically, 
and then comment.

The comparison uses the simplest case of a perfect linear atomic chain,
with 10-atom long leads with probes and a 4-atom central region without
probes. For simplicity we use a single-orbital orthogonal model, with
a nearest-neighbour hopping integral set to $-1$, defining the energy unit.
The corresponding energy band then lies in the energy interval $(-2,2)$,
and the 2-terminal Landauer solution has unit transmission 
throughout that interval.

\begin{figure}
\begin{centering}
\includegraphics[width=0.9\columnwidth]{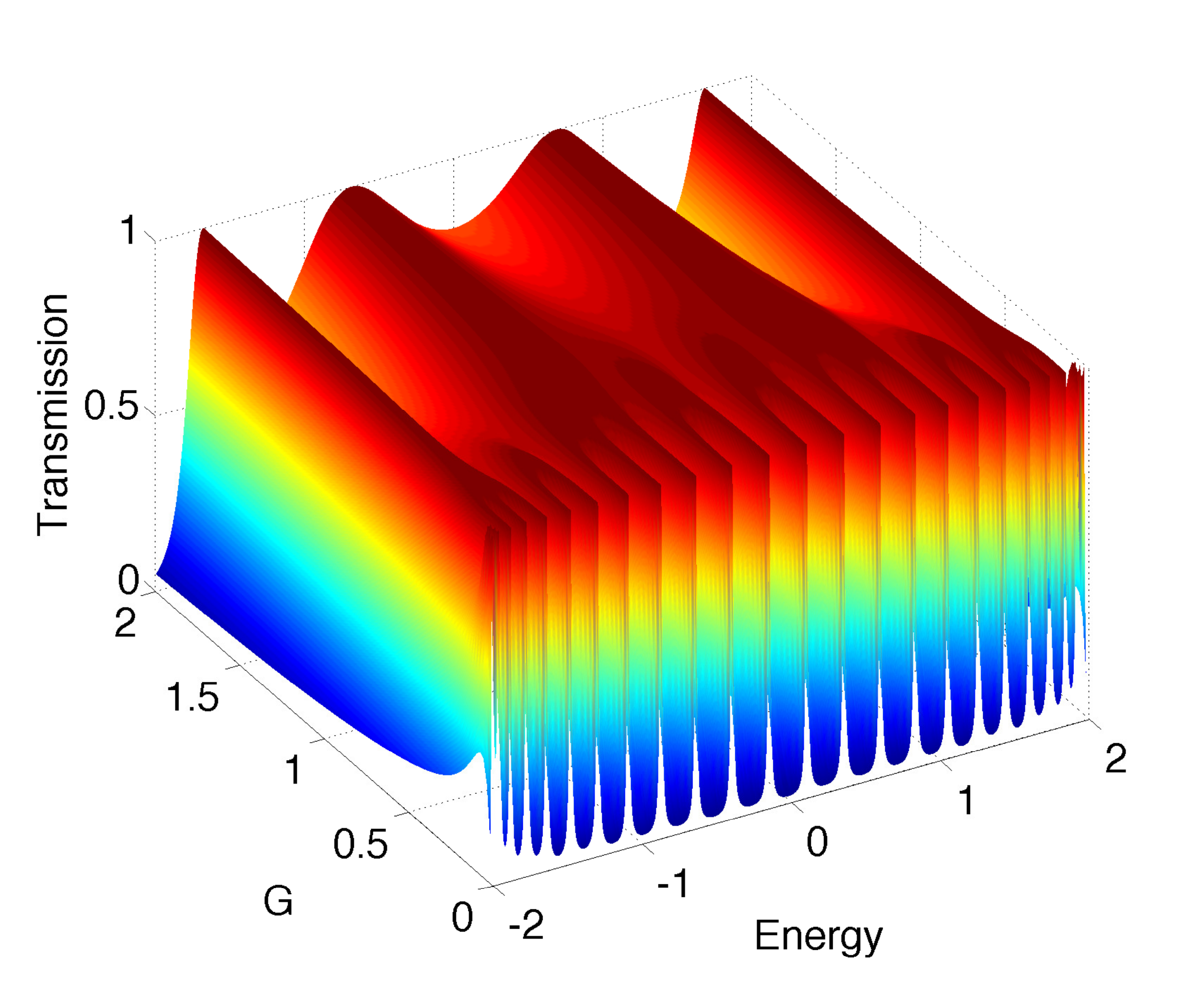}
\par\end{centering}
\caption{\label{fig:gradedA} The transmission as a function of energy
  for a perfect wire of length four atoms to which are connected two
  leads of length 10 atoms, each lead atom having one probe with a fixed coupling (G) attached.}
\end{figure}

First we consider the earlier implementation of the Hairy Probes,
with a position-independent coupling $\Gamma$ in each 10-atom
long lead. The surface plot in Fig. \ref{fig:gradedA} shows the transmission
as a function of energy and $\Gamma$.

Consider the limit of small $\Gamma$ first. 
In that limit, the $10+4+10$ atom system thinks of itself
as a 24-atom linear molecule weakly coupled to an envirnment, 
which just broadens its 24 molecular states into 24 narrow resonances. 
This is the origin of the 24 sharp transmission peaks at the small-$\Gamma$ 
end of the plot. To understand the opposite limit - large $\Gamma$ - consider 
first each 10-atom lead coupled to its probes, but not yet to the central
piece (corresponding to the isolated leads in the usual
Green's-function partitioned approach). If $\Gamma$ is big
enough, it dominates all other energy scales in the lead, ultimately 
making the lead itself a wide-band system, with a density of states 
(DOS) going down as $1/\Gamma$. If we now couple the components together, 
then the 4-atom central region just sees low-DOS adjoining leads, 
with a correspondingly small embedding self-energy. The upshot is that 
now the 4-atom central region behaves as a resonant system with weakly 
broadened states, producing the 4 resonances at the large-$\Gamma$ end 
of the plot.

In between these  two extremes, there is an optimal region of $\Gamma$-values, 
as expected, producing a roughly uniform transmission close to 1, but for the
given short leads the corrugation always remains visible. The reason is that
even at its optimal value, the finite $\Gamma$ results in an effective interface 
(between the regions with and without probes), which - like any interface - generates 
additional scattering. The longer the leads - and the smaller the optimal 
$\Gamma$ - the weaker the disruption.

\begin{figure}
\begin{centering}
\includegraphics[width=0.9\columnwidth]{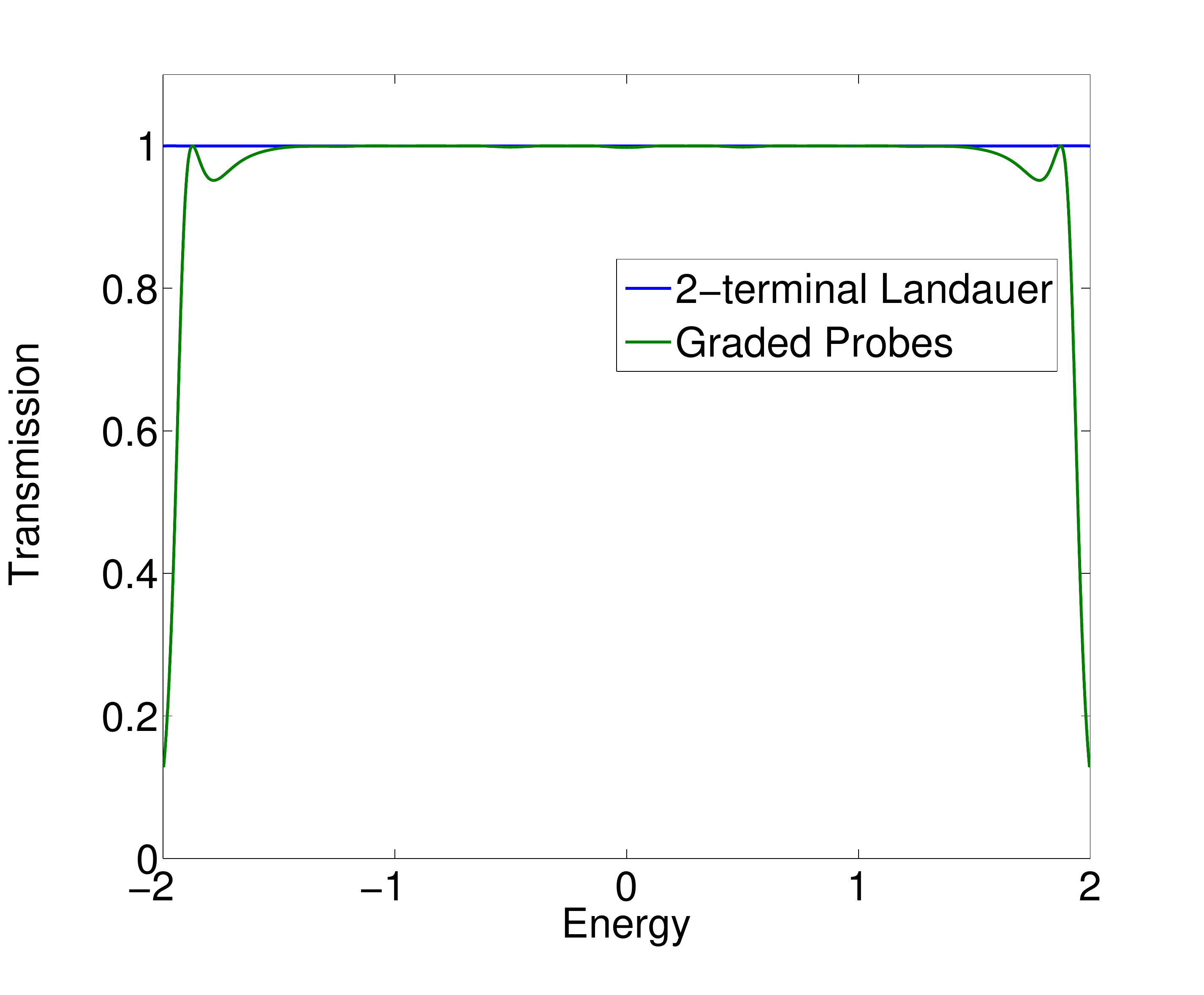}
\par\end{centering}
\caption{\label{fig:gradedB} The transmission as a function of energy
  for a perfect wire of length four atoms to which are connected two
  leads of length 10 atoms each having probes attached. The probes now
  have graded coupling strengths: see main text for the details.}
\end{figure}

The Graded Probes provide an alternative way to suppress this 
boundary scattering, without having to make the leads long. The plot in
Fig. \ref{fig:gradedB} shows the Graded Probes transmission, with $\Gamma$ rising linearly from 
zero to 1.4 along each 10-atom long lead. It is clear that - at no extra computational 
expense - we are now much closer to the ideal 2-terminal limit, even for the 
given modest lead length. The Graded Probes thus provide an alternative, 
if one wishes to avoid very long leads, or having to consider the precise choice 
of $\Gamma$ in the uniform-$\Gamma$ setup.

\section{Conclusions}

The primary purpose of this paper is to show how to extend the Hairy
Probe open boundary method for the steady state to non-orthogonal
atomic orbital basis sets. By considering the well understood case
of the one dimensional atomic wire (using both orthogonal and non-orthogonal
basis sets) we find that we obtain the expected conductance provided
the coupling of the wire to the probes has a suitable value. Couplings
that are either too large or too small suppress the current: small
couplings reduce the rate of charge injection, while large couplings
result in high levels of reflection of electrons back into the probes.
The optimal value results in the broadening of the system states by
the probes to produce a continuous density of states. There is still
more work to be done to understand completely the properties of the
probes. In addition to studying the dependence of current on applied
bias, there are a number of calculations that could be performed,
such as the transmission as a function of electron energy for different
couplings, or the self-consistent charge distribution.

The method is sufficiently simple that it can be implemented by finding
the eigenvalues and eigenvectors of an effective energy independent
Hamiltonian, and then performing all the subsequent integrals over
energy analytically to produce the single particle density matrix.
This results in an efficient algorithm that makes self-consistent
open boundary simulations easy to carry out, as it eliminates the
need to construct lead self-energies and to perform numerical integrals
over energy. The most time consuming part of the calculations is the
construction of the density matrices (Eq. \ref{eq:formalism-06e}).
For sparse Hamiltonian and overlap matrices, the scaling for building
the density matrix is $O(N^{3})$, which is no worse than the diagonalization
step. The absence of numerical integrals also helps keep the prefactor
low.

The method was applied to the problem of current flow through a benzene
ring attached by covalent bonds to two graphene contacts. It was found
that the formation of the contact covalent bonds disrupts the $\pi$-conjugation,
and thus heavily suppresses the current. We thus conclude that the
contacts must either involve physisorption, or a different way to
form covalent bonds must be found.

We have also introduced a possible way to accelerate the convergence
of the current with respect to lead length by using graded coupling
strengths for the probes. The results shown here look very promising,
though more work is needed to fully undeerstand them.

\begin{acknowledgments}
We gratefully acknowledge funding from the Leverhulme Trust (RPG-2014-125).
M.B. was supported by funding from EOARD (FA8655-12-1-2105) and through
a studentship in the Centre for Doctoral Training on Theory and Simulation
of Materials at Imperial College London, funded by EPSRC under Grant
No. EP/G036888/1. A.T. was funded by A-star. We also gratefully acknowledge
support from the Thomas Young Centre under grant TYC-10. R.D\textquoteright A.
acknowledges support by NANOTherm (CSD2010-00044) of the Spanish Ministerio
de Economia y Competitividad, and the Grupo Consolidado UPV/EHU del
Gobierno Vasco (Grant No. IT578-13). Finally we thank Matas Petreikis
for providing improved data for Fig. 4b.
\end{acknowledgments}

\appendix
\section{Analytic integrals over energy}

For electrons at finite temperature, in principle we should use the
Fermi-Dirac distribution, $f^{(p)}(E)=\left[1+\mathrm{e}^{(E-\mu_{p})/k_{B}T}\right]^{-1}$.
However, it is not then possible to evaluate the integrals analytically.
We thus use the piecewise linear approximation from
Eq. \ref{eq:formalism-07} with $T_p =T$ for all $p$:
\begin{equation}
\tilde{f}^{(p)}(E)=\begin{cases}
1 & E-\mu_{p}\le-2k_{B}T\\
\frac{1}{2}-\frac{1}{4}\left(\frac{E-\mu_{p}}{k_{B}T}\right) & -2k_{B}T<E-\mu_{p}<+2k_{B}T\\
0 & E-\mu_{p}\ge+2k_{B}T
\end{cases}\label{eq:HP-FT-01}
\end{equation}
If we define
\begin{eqnarray}
\mathcal{J}_{p}(\epsilon) & = & \left(\frac{\mu_{p}+2k_{B}T-\epsilon}{4k_{B}T}\right)\ln\left(\mu_{p}+2k_{B}T-\epsilon\right)\nonumber \\
 &  & -\left(\frac{\mu_{p}-2k_{B}T-\epsilon}{4k_{B}T}\right)\ln\left(\mu_{p}-2k_{B}T-\epsilon\right)\nonumber \\
 &  & -1-\ln\left(E_{pc}-\epsilon\right)\label{eq:HP-FT-01a}
\end{eqnarray}
then, using the piecewise linear approximation, the integrals become
\begin{eqnarray}
\mathcal{I}_{rs}^{(0)} & = & \int_{E_{p,c}}^{\infty}\frac{f^{(p)}(E)}{\left(E-\epsilon^{(r)}\right)\left(E-\epsilon^{(s)*}\right)}\,\mathrm{d}E\nonumber \\
 & = & \frac{1}{\epsilon^{(r)}-\epsilon^{(s)*}}\left\{ \mathcal{J}_{p}\left(\epsilon^{(r)}\right)-\mathcal{J}_{p}\left(\epsilon^{(s)*}\right)\right\} \label{eq:HP-FT-02a}\\
\mathcal{I}_{rs}^{(1)} & = & \int_{E_{p,c}}^{\infty}\frac{Ef^{(p)}(E)}{\left(E-\epsilon^{(r)}\right)\left(E-\epsilon^{(s)*}\right)}\,\mathrm{d}E\nonumber\\
 & = & \frac{1}{\epsilon^{(r)}-\epsilon^{(s)*}}\left\{ \epsilon^{(r)}\mathcal{J}_{p}\left(\epsilon^{(r)}\right)-\epsilon^{(s)*}\mathcal{J}_{p}\left(\epsilon^{(s)*}\right)\right\} \label{eq:HP-FT-02b}
\end{eqnarray}

\section{Formula for electric currents}

To evaluate the electric current that flows across a plane, we divide
our system into two parts (A and B), each defined by the list of atoms
within it. If we use an atomic orbital type basis set, this is equivalent
to defining the regions by the set of orbitals associated with the
atoms. We label orbitals in A by $\alpha$ and those in B by $\beta$.
The index for all orbitals (spanning A and B) shall be $\nu$. Let
the number of electrons in A be $N_{A}$, which can be computed from
the expression $N_{A}=2\mathrm{Tr}\left\{ \hat{\rho}\hat{P}_{A}\right\} $,
where $\hat{\rho}$ is the single particle electron density matrix
and $\hat{P}_{A}$ is a partition function for region A. We require
$\hat{P}_{A}$ to be symmetric and to satisfy $\hat{1}=\hat{P}_{A}+\hat{P}_{B}$,
where $\hat{P}_{B}$ is the corresponding partition function for region
B. Note that these partition functions are not projectors in general
as they need not be idempotent ($\hat{P}_{A}^{2}\ne\hat{P}_{A}$).
The current $\mathcal{I}_{A}$ is the time rate of change of the number
of electrons in $A$, namely
\begin{equation}
\mathcal{I}_{A}=-e\frac{\mathrm{d}N_{A}}{\mathrm{d}t}=-\frac{2e}{\mathrm{i}\hbar}\mathrm{Tr}\left\{ \hat{\rho}\left[\hat{P}_{A},\hat{H}\right]\right\} \label{eq:curr-03}
\end{equation}
where we have used the quantum Liouville equation, the fact that operators
permute under a trace, and $\hat{H}$ is the Hamiltonian. Note that
the matrix of coefficients of the density matrix is defined by $\hat{\rho}=\sum_{\nu\nu'}\left|\nu\right\rangle \rho_{\nu\nu'}\left\langle \nu'\right|$,
and the inverse overlap matrix we call $T(=S^{-1})$. We now define
$\hat{P}_{A}=\sum_{\nu\nu'}\left|\nu\right\rangle P_{A,\nu\nu'}\left\langle \nu'\right|$
and let the matrix of coefficients $P_{A}$ have the form
\begin{equation}
P_{A}=\left(\begin{array}{cc}
T_{AA} & \frac{1}{2}T_{AB}\\
\frac{1}{2}T_{BA} & 0
\end{array}\right)\label{eq:curr-08}
\end{equation}
where $S_{\alpha\alpha'}=(S_{AA})_{\alpha\alpha'}$ etc. The current
is then found to be
\begin{equation}
\mathcal{I}_{A}=-\frac{4e}{\hbar}\sum_{\alpha\beta}\left(H_{\beta\alpha}\mathrm{Im}\rho_{\beta\alpha}-S_{\beta\alpha}\mathrm{Im}E_{\beta\alpha}\right)\label{eq:curr-09}
\end{equation}
where the matrix $E$ satisfies $H\rho=SE$, and we have made use
of the fact that $H$ and $S$ are symmetric, while $\rho$ and $E$
are Hermitian. We can interpret Eq. \ref{eq:curr-09} as a sum over
bond currents, $\mathcal{I}_{A}=\sum_{\alpha\beta}\mathcal{I}_{\alpha\beta}$,
where $\mathcal{I}_{\alpha\beta}=-\frac{4e}{\hbar}\left(H_{\beta\alpha}\mathrm{Im}\rho_{\beta\alpha}-S_{\beta\alpha}\mathrm{Im}E_{\beta\alpha}\right)$.

\section{Choice of $E_{pc}$}

In the Hairy Probe formalism we assume the self-energies are independent
of energy; this allows us to use the simple spectral representation
of the Green's function. Having results depend on the value of the
lower cutoff in the integrals is not consistent with this assumption.
Here we investigate the internal consistency of the theory.

Let us rewrite Eq. \ref{eq:HP-FT-01a} as a sum of a term that is
independent of $E_{pc}$ ($\mathcal{K}_{p}(\epsilon)$) and a term
that depends on $E_{pc}$
\begin{equation}
\mathcal{J}_{p}(\epsilon)=\mathcal{K}_{p}(\epsilon)-\ln\left(E_{pc}-\epsilon\right)\label{eq:HP-FT-06}
\end{equation}
where
\begin{eqnarray}
\mathcal{K}_{p}(\epsilon) & = & \left(\frac{\mu_{p}+2k_{B}T-\epsilon}{4k_{B}T}\right)\ln\left(\mu_{p}+2k_{B}T-\epsilon\right)-1\nonumber \\
 &  & -\left(\frac{\mu_{p}-2k_{B}T-\epsilon}{4k_{B}T}\right)\ln\left(\mu_{p}-2k_{B}T-\epsilon\right)\label{eq:HP-FT-07}
\end{eqnarray}
Substituting Eq. \ref{eq:HP-FT-06} into Eqs. \ref{eq:HP-FT-02a} and \ref{eq:HP-FT-02a} then
gives
\begin{eqnarray}
\mathcal{I}_{rs}^{(0)} & = & \frac{\mathcal{K}_{p}\left(\epsilon^{(r)}\right)-\mathcal{K}_{p}\left(\epsilon^{(s)*}\right)}{\epsilon^{(r)}-\epsilon^{(s)*}}\nonumber \\
 &  & -\frac{1}{\epsilon^{(r)}-\epsilon^{(s)*}}\ln\left(\mathrm{sign}(E_{pc})-\frac{\epsilon^{(r)}}{\left|E_{pc}\right|}\right)\nonumber \\
 &  & +\frac{1}{\epsilon^{(r)}-\epsilon^{(s)*}}\ln\left(\mathrm{sign}(E_{pc})-\frac{\epsilon^{(s)*}}{\left|E_{pc}\right|}\right)\nonumber \\
\mathcal{I}_{rs}^{(1)} & = & \frac{\epsilon^{(r)}\mathcal{K}_{p}\left(\epsilon^{(r)}\right)-\epsilon^{(s)*}\mathcal{K}_{p}\left(\epsilon^{(s)*}\right)}{\epsilon^{(r)}-\epsilon^{(s)*}}-\ln\left|E_{pc}\right|\nonumber \\
 &  & -\frac{\epsilon^{(r)}}{\epsilon^{(r)}-\epsilon^{(s)*}}\ln\left(\mathrm{sign}(E_{pc})-\frac{\epsilon^{(r)}}{\left|E_{pc}\right|}\right)\nonumber \\
 &  & +\frac{\epsilon^{(s)*}}{\epsilon^{(r)}-\epsilon^{(s)*}}\ln\left(\mathrm{sign}(E_{pc})-\frac{\epsilon^{(s)*}}{\left|E_{pc}\right|}\right)\label{eq:HP-FT-08}
\end{eqnarray}
In the limit that $\left|E_{pc}\right|\gg\left|\epsilon\right|$,
and for $E_{pc}<0$, we have
\begin{eqnarray}
\ln\left(\mathrm{sign}(E_{pc})-\frac{\epsilon}{\left|E_{pc}\right|}\right) & = & \ln\left|\mathrm{sign}(E_{pc})-\frac{\epsilon}{\left|E_{pc}\right|}\right|\nonumber \\
 &  & +\mathrm{i}\arg\left(\mathrm{sign}(E_{pc})-\frac{\epsilon}{\left|E_{pc}\right|}\right)\nonumber \\
 & \to & -\mathrm{i}\pi\mathrm{sign}(\Im\epsilon)\label{eq:HP-FT-08a}
\end{eqnarray}
Substituting Eq. \ref{eq:HP-FT-08a} into Eq. \ref{eq:HP-FT-08},
and noting that $\mathrm{sign}(\Im\epsilon^{(r)})=-1$, we get

\begin{eqnarray}
\mathcal{I}_{rs}^{(0)} & \to & \frac{\mathcal{K}_{p}\left(\epsilon^{(r)}\right)-\mathcal{K}_{p}\left(\epsilon^{(s)*}\right)-\mathrm{i}2\pi}{\epsilon^{(r)}-\epsilon^{(s)*}}\nonumber\\
\mathcal{I}_{rs}^{(1)} & \to &
                               \frac{\epsilon^{(r)}\mathcal{K}_{p}\left(\epsilon^{(r)}\right)-\epsilon^{(s)*}\mathcal{K}_{p}\left(\epsilon^{(s)*}\right)-\mathrm{i}\pi\left(\epsilon^{(r)}+\epsilon^{(s)*}\right)}{\epsilon^{(r)}-\epsilon^{(s)*}}\nonumber\\
&&-\ln\left|E_{pc}\right|\label{eq:HP-FT-09}
\end{eqnarray}
Thus for large enough $\left|E_{pc}\right|$, $\mathcal{I}_{rs}^{(0)}$
becomes independent of $E_{pc}$, while $\mathcal{I}_{rs}^{(1)}$
varies with $E_{pc}$ as $-\ln\left|E_{pc}\right|$, which is independent
of $r$ and $s$. We now recall the expressions for the density matrices
\begin{eqnarray}
\rho_{\beta\beta'} & = & \frac{1}{2\pi}\sum_{p}\Gamma_{p}\sum_{rs}\mathcal{I}_{rs}^{(0)}\zeta_{\beta_{p}}^{(r)*}\chi_{\beta}^{(r)}\zeta_{\beta_{p}}^{(s)}\chi_{\beta'}^{(s)*}\nonumber \\
E_{\beta\beta'} & = & \frac{1}{2\pi}\sum_{p}\Gamma_{p}\sum_{rs}\mathcal{I}_{rs}^{(1)}\zeta_{\beta_{p}}^{(r)*}\chi_{\beta}^{(r)}\zeta_{\beta_{p}}^{(s)}\chi_{\beta'}^{(s)*}\label{eq:HP-FT-10}
\end{eqnarray}
Let us define $\Delta\rho_{\beta\beta'}(E_{pc})$ and $\Delta E_{\beta\beta'}(E_{pc})$
to be those parts of the density matrices that depend on $E_{pc}$.
Combining Eq. \ref{eq:HP-FT-09} with Eq. \ref{eq:HP-FT-10} we get
\begin{eqnarray}
\Delta\rho_{\beta\beta'}(E_{pc}) & \to & \frac{1}{2\pi E_{pc}}\sum_{p}\Gamma_{p}\sum_{r}\zeta_{\beta_{p}}^{(r)*}\chi_{\beta}^{(r)}\sum_{s}\zeta_{\beta_{p}}^{(s)}\chi_{\beta'}^{(s)*}\nonumber \\
 & = & \frac{1}{2\pi E_{pc}}\sum_{p}S_{\beta\beta_{p}}^{-1}\Gamma_{p}S_{\beta_{p}\beta'}^{-1}\label{eq:HP-FT-11a}\\
\Delta E_{\beta\beta'}(E_{pc}) & \to & -\frac{1}{2\pi}\ln\left|E_{pc}\right|\sum_{p}\Gamma_{p}\sum_{r}\zeta_{\beta_{p}}^{(r)*}\chi_{\beta}^{(r)}\sum_{s}\zeta_{\beta_{p}}^{(s)}\chi_{\beta'}^{(s)*}\nonumber \\
 & = & -\frac{1}{2\pi}\ln\left|E_{pc}\right|\sum_{p}S_{\beta\beta_{p}}^{-1}\Gamma_{p}S_{\beta_{p}\beta'}^{-1}\label{eq:HP-FT-11b}
\end{eqnarray}
The contribution from $\Delta\rho_{\beta\beta'}$ becomes arbitrarily
small for large enough $E_{pc}$, while the contribution from $\Delta E_{\beta\beta'}$
is logarithmically divergent. We note that the $E_{pc}$ dependent
parts of both matrices are real and symmetric, thus they make no contribution
to the electric current, but make a contribution to the atomic forces
(not discussed further in this paper). These contributions decrease
as the distance from the probes increases, and can be suppressed entirely
if we set to zero those overlap matrix elements that link orbitals
not attached to probes to those that are attached to probes.

In the main text we offer an alternative interpretation of $E_{pc}$
that allows us to avoid these difficulties, but at the expense of
being unphysical: it can be interpreted as the lowest energy for which
states in the probes are populated.


%

\end{document}